\documentclass[12pt,a4paper,onecolumn,prb,showpacs]{revtex4}

\usepackage[pdfborder={0 0 0},colorlinks=true]{hyperref}

\usepackage{epsfig}
\usepackage{amsmath}
\usepackage{amstext}
\usepackage{amssymb}
\usepackage{amsfonts}
\usepackage{graphicx}
\usepackage{color}

\def\beq{\begin{equation}}
\def\eeq{\end{equation}}
\def\bea{\begin{eqnarray}}
\def\eea{\end{eqnarray}}

\newcommand{\p}{^{\prime}}

\def\q{\mathbf{q}}
\def\k{\mathbf{k}}

\def\p{\mathbf{p}}

\def\Q{\mathbf{Q}}

\newcommand{\eps}{\varepsilon}

\newcommand{\su}{\uparrow}
\newcommand{\sd}{\downarrow}
\newcommand{\sgn}[1] {\mathrm{sgn}\left({#1}\right)}

\newcommand{\ii}{{\mathrm{i}}}
\newcommand{\vk}{{\mathbf{k}}}

\newcommand{\nn}{\nonumber}

\newcommand{\half}{\frac{1}{2}}

\newcommand{\ppm}{{+-}}

\newcommand{\av}[1]{\left<{#1}\right>}

\newcommand{\ket}[1]{\left|{#1}\right>}

\begin{document}


\title{Superconducting state in Fe-based materials and spin-fluctuation theory of pairing
\footnote{Published in\\ \href{http://dx.doi.org/10.3367/UFNr.0184.201408h.0882}{\textit{Uspekhi Fizicheskikh Nauk} \textbf{184} (8), 882-888 (2014)}, DOI: 10.3367/UFNr.0184.201408h.0882 (in Russian)\\
\href{http://dx.doi.org/10.3367/UFNe.0184.201408h.0882}{\textit{Physics-Uspekhi} \textbf{57} (8), 813-819 (2014)}, DOI: 10.3367/UFNe.0184.201408h.0882 (in English)\\
\textit{Translated by S.N. Gorin; edited by A. Radzig.}}
}
\author{M.M. Korshunov}
\affiliation{L.V. Kirensky Institute of Physics SB RAS, 660036, Krasnoyarsk, Russia\\
e-mail: mkor@iph.krasn.ru}

\date{\today}

\maketitle

\begin{flushright}
\scriptsize
\item[] \textit{
Beyond the pairs of opposites of which the world consists, other, new insights begin.\\
Herman Hesse, ``Inside and Outside'', in Stories of
Five Decades (London: Jonathan Cape, 1974)}
\end{flushright}

Quite recently, the scientific community was shaken up by a new discovery. In the field of high-$T_c$ superconductivity, where cuprates had overwhelmingly predominated for the
previous two decades, a new player -- iron-based materials -- has appeared~\cite{Kamihara}. Although the superconducting transition temperature ($T_c$) in iron-based compounds has not exceeded the liquid-nitrogen temperature, already in late 2008, i.e., less than a year after the discovery of this new class of superconducting materials, this temperature reached 56K. To date, the record among single crystals has belonged to SmFeAsO$_{1-x}$F$_x$ ($T_c = 57.5$K)~\cite{SmFeAsOTc}; great hopes have been laid upon the discovery of superconductivity with $T_c \sim 60$K in single-layer FeSe films~\cite{FeSeTc,FeSeARPES}.

In general, superconducting iron materials can be grouped in two classes: pnictides, and chalcogenides. The basic element in these compounds is a square lattice of iron atoms, which in the majority of weakly doped compounds is subjected to orthorhombic distortions at temperatures comparable with the temperature of the transition to the antiferromagnetic (AFM) phase, $T_{SDW}$.  In compounds of the first class, iron resides in a tetrahedral surrounding of arsenic or phosphorus atoms; in compounds of the second
class, of selenium, tellurium, or sulfur atoms. The pnictides can be single-layer, e.g., 1111 (LaFeAsO, LaFePO, Sr$_2$VO$_3$FeAs, etc.) and 111 materials (LiFeAs, LiFeP, etc.),  and two-layer 122 materials with two layers of FeAs per unit cell (BaFe$_2$As$_2$, KFe$_2$As$_2$, etc.). The chalcogenides include compounds of the 11 type (Fe$_{1-\delta}$Se, Fe$_{1+y}$Te$_{1-x}$Se$_x$, FeSe films) and of the 122 type (KFe$_2$Se$_2$). The structure and physical properties of iron compounds have been discussed in detail in many reviews (see, e.g. Refs.~\onlinecite{SadovskiiReview,IvanovskiiReview,IzyumovReview,IshidaReview,JohnstonReview,PaglioneReview,MazinReview,LumsdenReview,WenReview,BasovReview,StewartReview,review}).

A characteristic feature of iron compounds as opposed to, for instance, cuprates, consists in a qualitative, or sometimes even quantitative, agreement of their Fermi surface (measured by the angle-resolved photoemission spectroscopy (ARPES) and by the quantum oscillations) with the Fermi surface calculated from the first principles. This peculiarity, together with the small magnitude of the magnetic moment of iron atoms ($\sim 0.3 \mu_B$) in the pnictides and the absence of the dielectric state in the undoped case, make it possible to speak of a small or moderate level of electron correlations. Therefore, the natural starting point for their description is the model of itinerant electrons rather than the Mott-Hubbard limit and $t-J_1-J_2$ type models.

Soon after the discovery of superconductivity in pnictides, estimates were made of the possibility of pairing due to the electron-phonon interaction. The coupling constant appears to be even smaller than that for aluminum~\cite{Boeri}, although $T_c$ c in
iron compounds is significantly higher. This led to the conclusion that it is unlikely that the pairing caused by electron-phonon interaction could play a leading role in the
emergence of superconductivity, although a more thorough analysis is probably required to take into account some specific features of the electron band structure~\cite{Eschrig}. Such a situation immediately led to searching for alternative theories of superconducting pairing. The interactions that are analyzed in the theories vary from spin and orbital fluctuations to strongly correlated Hubbard and Hund's exchanges. It is unrealistic to describe or even simply mention all these theories in the present paper; therefore, we focus on one of the most promising theories, namely, the spin-fluctuation theory of the superconducting pairing.

The spin-fluctuation theory of superconductivity is promising for a number of reasons: (1) this theory is based on the model of itinerant electrons, which serves as a good starting point for the description of iron compounds; (2) the superconducting phase arises directly after the AFM phase or coexists with it; in this case, the character of the spin-lattice relaxation rate $1/T_1T$  gradually changes from the Curie-Weiss to Pauli behaviors with an increase of doping and decreasing $T_c$~\cite{Ning},  which indicates a decrease in the role of spin fluctuations; (3) the description of various experimentally observed properties of the pnictides and chalcogenides does not require the introduction of additional parameters in to the theory; rather, only some specific features of the band structure and of the interactions in different classes of the iron compounds should be taken into account~\cite{review}.

\begin{figure}[t]
\includegraphics[width=0.5\textwidth]{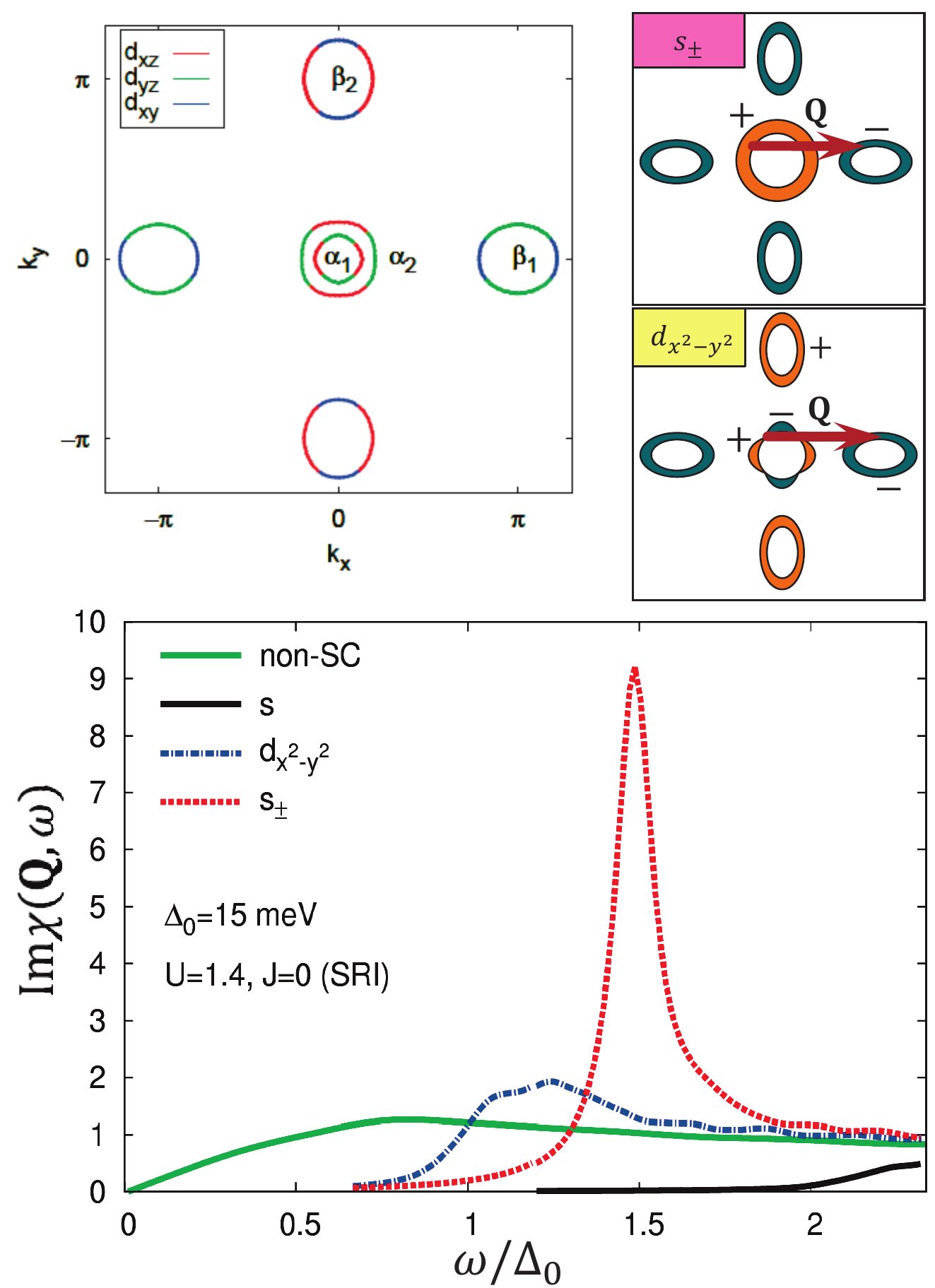}
\caption{\label{fig:spinres} (Color online) Fermi surface (top left) in the model of Ref.~\onlinecite{Graser} upon electron doping $x=0.05$ in the Brillouin zone corresponding to one Fe atom per unit cell. The orbitals making the maximum contributions to the Fermi surface are shown by different colors. Frequency dependence of the susceptibility $\mathrm{Im}\chi(\q=\Q,\omega)$  in the normal state (non-SC) and in the superconducting state with the $s$ ($s_{++}$), $d_{x^2-y^2}$, and $s_\pm$ symmetries of the gap (bottom). The calculation was done for Hubbard interaction $U=1.4$eV and Hund's exchange $J=0$ in the presence of the spin-rotational invariance (SRI). For the $s_\pm$ state, a resonance peak appears for $\omega < 2 \Delta_0$. Schematic structures of the $s_\pm$ and $d_{x^2-y^2}$ superconducting gaps on the Fermi surface are shown on the top right panels; vector $\Q=(\pi,0)$ connects electron and hole pockets.}
\end{figure}

Iron-based superconductors represent quasi-two-dimensional substances in which the square lattice of iron atoms serves as the conducting plane. As was shown by the early calculations in the density functional theory (DFT)~\cite{Lebegue,Singh,Mazin2008}, which satisfactorily agree with the results of ARPES and quantum oscillation measurements, it is the $3d^6$ states of Fe$^{2+}$ that are dominant near the Fermi level. Moreover, all five $d$-orbitals ($d_{x^2-y^2}$, $d_{3z^2-r^2}$, $d_{xy}$, $d_{xz}$, and $d_{yz}$) lie on or near the Fermi surface. This leads to the substantial multiorbital and multiband character of the low-energy electron structure, which cannot be described already in a one-band model. Thus, for example, in the five-orbital model from Ref.~\onlinecite{Graser}, which satisfactorily describes the results of DFT calculations~\cite{Cao}, the Fermi surface consists of four pockets: two hole pockets near the $(0,0)$ point and two electron pockets in the vicinity of the $(\pi,0)$ and $(0,\pi)$ points (see Fig.~\ref{fig:spinres}). Such geometry in the $\k$-space leads to the possibility of generating the spin density wave (SDW) order because of the nesting between the
hole and electron Fermi surfaces with the wave vector $\Q=(\pi,0)$ or $\Q=(0,\pi)$. With increasing doping $x$,  the long-range SDW order disappears. In the case of electron doping, the hole pockets disappear when $x$ exceeds some particular value, and only electron pockets are retained, as, e.g., in the case of K$_x$Fe$_{2-y}$Se$_2$ and FeSe monolayers~\cite{FeSeARPES}. With increasing hole concentration, a new hole pocket appears around the $(\pi,\pi)$ point and then the electron Fermi surfaces disappear. This situation arises, in particular, in KFe$_2$As$_2$. The fact that the maximum contributions to the band lying on the Fermi surface come from the $d_{xz,yz}$ and $d_{xy}$ orbitals is confirmed by ARPES~\cite{Kordyuk,Brouet}. In this case, as will be shown below, the existence of several electron pockets and the multiorbital character of the bands substantially affect the superconducting pairing.

\begin{figure}[t]
\includegraphics[width=\textwidth]{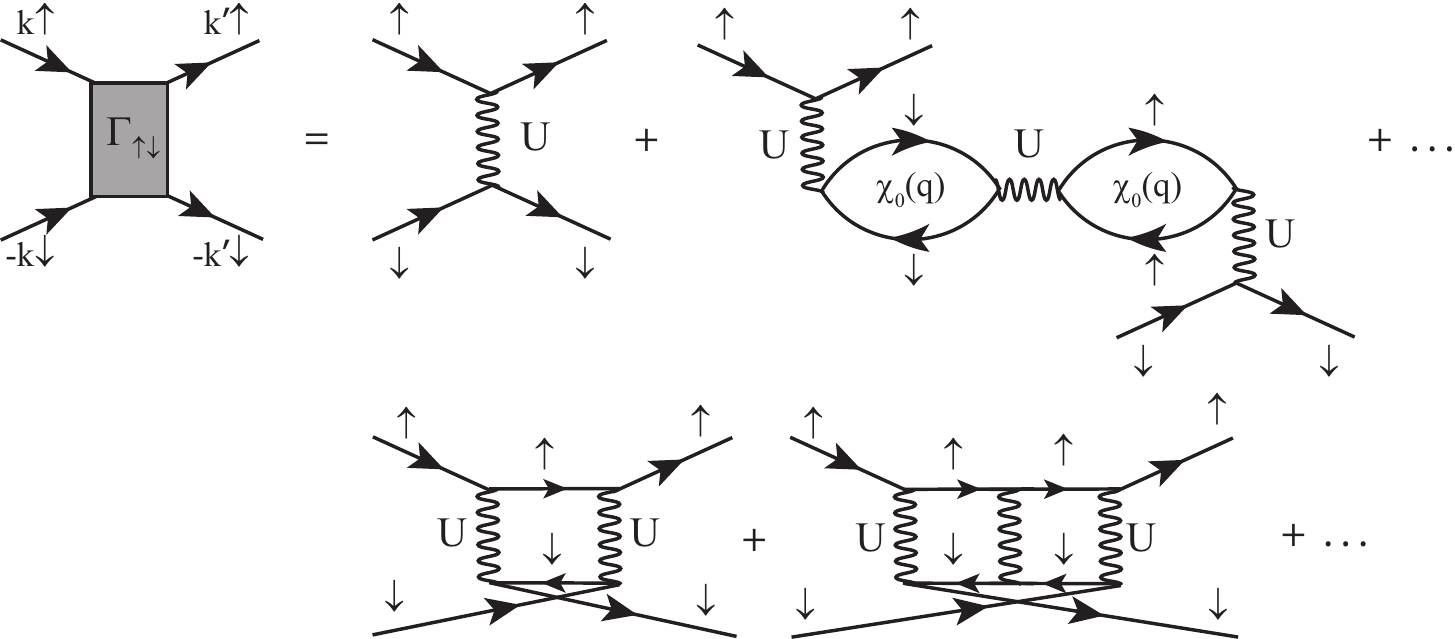}
\caption{\label{fig:sfpairing} Cooper vertex $\Gamma_{\su\sd}$ for a singlet superconducting state in the RPA.}
\end{figure}

Before moving to the description of the multiorbital variant of the theory, let us describe how the spin-fluctuation theory of pairing is constructed in the single-band case with a Hubbard interaction Hamiltonian $H = \sum_f U n_{f\su} n_{f\sd}$, where $U$ is the single-site Coulomb (Hubbard) repulsion, and $n_{f\sigma}$ is the operator of the number of particles on the site $f$ with a spin $\sigma$. The superconducting interaction in the singlet channel is determined by the Cooper vertex $\Gamma_{\su\sd}$,  which, in the spirit of the Berk-Schrieffer theory~\cite{BerkSchrieffer,ScalapinoSF,ScalapinoHF}, is given by a diagrammatic series in the random-phase approximation (RPA) shown in Fig.~\ref{fig:sfpairing}. The basic element in this case is an electron-hole bubble, i.e., the `bare' susceptibility
\beq
\chi_0(\q,\ii\omega_n) = \sum_\p \frac{f(\eps_{\p+\q}) - f(\eps_{\p})}{\ii\omega_n - \eps_{\p+\q} + \eps_{\p}},
\eeq
where $f(\eps_{\p})$  is the Fermi distribution function for the electron
dispersion $\eps_{\p}$ and $\omega_n$ is the Matsubara frequency. The sum of
bubbles and ladders yields
\bea \label{eq:Gamma1band}
\Gamma_{\su\sd} &=& U (1 + U^2 \chi_0^2 + ...) + U^2 \chi_0 (1 + U \chi_0 + ... ) = \frac{U}{1 - U^2 \chi_0^2} + \frac{U^2 \chi_0}{1 - U \chi_0}\\
&=& \frac{3}{2} U^2 \chi_s - \frac{1}{2} U^2 \chi_c + U,
\eea
where $\chi_{s}$ and $\chi_{c}$ are the spin and charge susceptibilities, respectively:
\beq
\chi_{s,c} = \frac{\chi_0}{1 \mp U \chi_0}.
\eeq
A magnetic instability develops in the system if the Stoner criterion is fulfilled: $1 = U \chi_0(\q,\omega=0)$.  The ferromagnetic instability corresponds to $\q = 0$; the AFM instability, which we are interested in, appears at the antiferromagnetic wave
vector $\q = \Q$. If we avoid the development of the instability, for example, via doping, then no long-range order will appear, but the product $U \chi_0(\q,\omega=0)$  will be close to unity, thus leading to a large magnitude of the spin susceptibility $\chi_{s}$ and, correspondingly, to its very large contribution to the Cooper vertex $\Gamma_{\su\sd}$. However, unlike the electron-phonon attractive interaction in the Bardeen-Cooper-Schrieffer (BCS) theory, $\Gamma_{\su\sd}$ results in the effective \textit{repulsive} interaction $V(\k,\k')$. If we write the Hamiltonian of the system in terms of the mean-field theory, explicitly separating the superconducting interaction
\beq
H = \sum_{\k,\sigma} \eps_\k a_{\k\sigma}^\dag a_{\k\sigma} + \frac{1}{2} \sum_{\k,\k',\sigma} V(\k-\k') a_{-\k\sigma}^\dag a_{\k\bar{\sigma}}^\dag a_{\k'\sigma} a_{\k'\bar{\sigma}},
\eeq
where $\bar{\sigma} = -\sigma$ and $a_{\k\sigma}^\dag$ is the creation operator of an electron with a momentum $\k$ and spin $\sigma$, then the gap equation will take the form
\beq
\Delta_\k(T) = - \sum\limits_{\k'} \frac{V(\k-\k')}{2 E_{\k'}} \Delta_{\k'}(T) \tanh{\frac{E_{\k'}}{2T}},
\label{eq:delta}
\eeq
where $E_{\k} = \sqrt{\eps_{\k}^2 + \Delta_{\k}^2}$.  In the case of electron-phonon
interaction with a coupling constant $g_{e-ph}$ in the BCS theory, we have $V(\k-\k') = - g_{e-ph}^2$ and equation~(\ref{eq:delta}) has the solution $\Delta_\k = \Delta_0(T)$,  which corresponds to the $s$-type of the superconducting order parameter. In iron compounds, the orbital fluctuations enhanced by electron-phonon interaction can lead to a sign-constant solution, which in the multiband case is called the $s_{++}$ state~\cite{Kontani,Onari}.  On the other hand, for the spin-fluctuation interaction we have $V(\k-\k') > 0$ and the $s$-type of solution does not satisfy equation~(\ref{eq:delta}). In the case of spin fluctuations, $V(\k-\k')$  has a
maximum at the wave vector $\Q$, and if we use a very rough approximation, $V(\k-\k') = |\lambda| \delta(\k-\k'+\Q)$, then equation~(\ref{eq:delta}) will take the form
\beq
\Delta_\k(T) = - |\lambda| \frac{\Delta_{\k+\Q}(T)}{2 E_{\k+\Q}} \tanh{\frac{E_{\k+\Q}}{2T}}.
\eeq
It is obvious that the last equation has a solution if $\Delta_\k$ and $\Delta_{\k+\Q}$ have different signs. In the simplest case of $\Delta_\k = - \Delta_{\k+\Q}$ the equation acquires the form
\beq
1 = |\lambda| \frac{1}{2 E_{\k+\Q}} \tanh{\frac{E_{\k+\Q}}{2T}}.
\eeq
The solution defines a gap, which reverses sign at the vector $\Q$. If this vector connects different bands of the quasiparticles (Fermi surfaces belonging to different bands), which is realized, in particular, in iron-based materials, then the solution of this type with an $A_{1g}$ symmetry is called the $s_\pm$ state~\cite{Mazin2008}. The competing states will be those with a $B_{1g}$ and a $B_{2g}$ symmetries, namely, those that have the $d_{xy}$ and $d_{x^2-y^2}$ types of the order parameter.

In the multiorbital case, the central subject of the spin-fluctuation theory -- the dynamic spin susceptibility -- is a tensor with respect to the orbital indices $l$, $l'$, $m$, and $m'$:
\beq\label{eq.chi_def}
 \chi^{ll',mm'}_{ss'}(\q,\ii\Omega) = - \int\limits_0^\beta d\tau e^{\ii\Omega\tau} \av{T_\tau S^{s}_{ll'}(\q,\tau) S^{s'}_{m'm}(-\q,0)}.
\eeq
Here, $\Omega$ is the Matsubara frequency, $S^{s}_{ll'}(\q,\tau)$  is the $s$th component of the vector of the spin operator with the
Matsubara time $\tau$:
\beq
\vec{S}_{ll'}(\q,\tau) = \frac{1}{2} \sum_{\p,\alpha,\alpha'} a^\dag_{\p l \alpha}(\tau) \vec{\hat\sigma}_{\alpha\alpha'} a_{\p+\q l' \alpha'}(\tau),
\eeq
where $\vec{\hat\sigma}$ is a vector composed of Pauli matrices $\hat\sigma$, and $a^\dag_{\p l \alpha}$ is the operator of creation of an electron with the orbital index $l$, momentum $\p$, and spin $\alpha$. To obtain a zero's order approximation, we decouple expression~(\ref{eq.chi_def}) via Wick's theorem, introduce normal and anomalous Green's functions
\bea
G_{m l \sigma\sigma'}(k,\tau) &=& - \av{T_\tau a_{k m \sigma}(\tau) a^\dag_{k l \sigma'}(0)}, \nn\\
F^\dag_{m l \sigma\sigma'}(k,\tau) &=& \av{T_\tau a^\dag_{k m \sigma}(\tau) a^\dag_{-k l \bar\sigma'}(0)}, \nn\\
F_{m l \sigma\sigma'}(k,\tau) &=& \av{T_\tau a_{k m \sigma}(\tau) a_{-k l \bar\sigma'}(0)}, \nn
\eea
transform to the Matsubara frequencies $\omega_n$ and find the expression for the $\ppm$  component of the susceptibility in the absence of spin-orbit interaction:
\bea \label{eq.chipm}
 \chi^{ll',mm'}_{0,\ppm}(\q,\ii\Omega) = -T \sum_{\omega_n,\p} &&\left[ G_{m l \su}(\p,\ii\omega_n) G_{l' m' \sd\sd}(\p+\q,\ii\Omega+\ii\omega_n) \right.\nn\\
 &&\left. -F^\dag_{l m' \su\sd}(\p,-\ii\omega_n) F_{l' m \sd\su}(\p+\q,\ii\Omega+\ii\omega_n) \right].
\eea
The physical (observed) susceptibility is obtained at the coincident orbital indices of the two Green's functions entering into the vertex, i.e., at $l'=l$ and $m'=m$: $\chi_{+-}(\q,\ii\Omega) = \half \sum\limits_{l,m} \chi^{ll,mm}_{+-}(\q,\ii\Omega)$.

The Cooper vertex $\Gamma_{\su\sd}$ has to be calculated in the normal phase, where there are no anomalous Green's functions. The Green's functions in the orbital basis are off-diagonal and depend on two orbital indices. It makes sense to move to the band basis constructed using operators of electron creation and annihilation, $b_{\k \mu \sigma}^\dag$ and $b_{\k \mu \sigma}$ with the band index $\mu$,  where
the Green's functions are diagonal: $G_{\mu \sigma}(\k,\ii\Omega) = 1 / \left( \ii\Omega - \eps_{\k\mu\sigma} \right)$. The transition from the orbital to band basis is implemented with the aid of the matrix elements $\varphi^{\mu}_{\k m}$: $\ket{\sigma m \vk} = \sum\limits_{\mu} \varphi^{\mu}_{\vk m} \ket{\sigma \mu \vk}$. In this case, $a_{\k m \sigma} = \sum\limits_{\mu} \varphi^{\mu}_{\k m} b_{\k \mu \sigma}$ and the
susceptibility takes the form
\beq \label{eq.chipmmu}
 \chi^{ll',mm'}_{0,+-}(\q,\ii\Omega) = -T \sum_{\omega_n, \p, \mu,\nu} \varphi^{\mu}_{\p m} {\varphi^*}^{\mu}_{\p l} G_{\mu \su}(\p,\ii\omega_n) G_{\nu \sd}(\p+\q,\ii\Omega+\ii\omega_n) \varphi^{\nu}_{\p+\q l'} {\varphi^*}^{\nu s'}_{\p+\q m'}.
\eeq

Below, we will rely on the model of the band structure $H_0$ from Ref.~\onlinecite{Graser}, which is based on DFT calculations~\cite{Cao} for a single-layer LaFeAsO pnictide. As the interaction, we will use the two-particle Hamiltonian with a single-site interaction of the general form~\cite{Castellani,Oles,Kuroki2008,Graser}:
\bea
H &=& H_{0} + U \sum_{f, m} n_{f m \su} n_{f m \sd} + U' \sum_{f, m < l} n_{f l} n_{f m} \nn\\
  && + J \sum_{f, m < l} \sum_{\sigma,\sigma'} a_{f l \sigma}^\dag a_{f m \sigma'}^\dag a_{f l \sigma'} a_{f m \sigma}
  + J' \sum_{f, m \neq l} a_{f l \su}^\dag a_{f l \sd}^\dag a_{f m \sd} a_{f m \su}.
\label{eq:H}
\eea
where $n_{f m} = n_{f m \su} + n_{f m \sd}$, $f$ is the site index, $U$ and $U'$ are
the intra- and interorbital Hubbard repulsions, $J$ is Hund's exchange, and $J'$ is the pair hopping. The parameters usually obey the spin-rotational invariance (SRI), which leads to a decrease in the number of free parameters of the theory because of the relationships  $U' = U - 2J$ and $J' = J$.

Based on the interaction in Hamiltonian~(\ref{eq:H}) we can formulate the RPA for the spin susceptibility $\chi_{+-}(\q,\ii\Omega)$~\cite{Graser}. To obtain the solution, we transform from tensors to matrices with the indices $\imath = l + l' n_O$ and $\jmath = m + m' n_O$, where $n_O$ is the number of orbitals. Then, in the matrix form, the spin susceptibility in the RPA is written down as
\beq \label{eq.rpa.chippmsol}
 \hat\chi_\ppm = \left( \hat{1} - \hat\chi_{0,\ppm} \hat{U}^\ppm \right)^{-1} \hat\chi_{0,\ppm},
\eeq
where $\hat{U}^\ppm$ is the interaction matrix in the $\ppm$ channel.

The Cooper vertex in the multiorbital case is similar to that in the single-band case~(\ref{eq:Gamma1band}),
\beq \label{eq:GammaOrb}
\Gamma_{\su\sd}^{l_1 l_2 l_3 l_4}(\k,\k',\omega) = \left[ \frac{3}{2} \hat{U_s} \hat\chi_s(\k-\k',\omega) \hat{U_s} - \frac{1}{2} \hat{U_c} \hat\chi_c(\k-\k',\omega) \hat{U_c} + \frac{1}{2} \hat{U_s} + \frac{1}{2} \hat{U_c} \right]_{l_1 l_2 l_3 l_4},
\eeq
where $\hat\chi_{s,c} = \left( \hat{1} \mp \hat\chi_{0} \hat{U}_{s,c} \right)^{-1} \hat\chi_{0}$ is the spin ($s$) and charge ($c$) susceptibilities, $\hat{U}_{s,c}$ are the interaction matrices in the spin and charge channels, and $l_1$ to $l_4$ are the orbital indices.

The necessity of constructing the theory in orbital representation stems from the fact that just in this representation the Hubbard interaction~(\ref{eq:H}), remains local. The superconducting pairs, however, are formed at the level of bands rather than orbitals; therefore, we should transform the Cooper vertex into a band basis via matrix elements $\varphi^{\mu}_{\k m}$,
\beq \label{eq:GammaBand}
\Gamma^{\mu\nu}(\k,\k',\omega) = \sum\limits_{l_1,l_2,l_3,l_4} \varphi^{\mu*}_{\k l_2} \varphi^{\mu*}_{-\k l_3} \Gamma_{\su\sd}^{l_1 l_2 l_3 l_4}(\k,\k',\omega) \varphi^{\nu}_{\k' l_1} \varphi^{\nu}_{-\k' l_4}.
\eeq
Calculations show that $\Gamma^{\mu\nu}$ rapidly decreases with increasing frequency in the range of frequencies that are much lower than the band width. Although the equation for the superconducting gap depends, generally speaking, on $\mathrm{Im}\Gamma^{\mu\nu}$, the momenta $\k$ and $\k'$ making the main contribution to the pairing should correspond to the small frequencies at which these momenta lie near the Fermi surface. Similarly to the case where the coupling constant for the electron-phonon interaction is determined by the integral of the Eliashberg function $\alpha^2 F(\omega)$ taken with respect to frequency, here, using the Kramers-Kronig relationship, we obtain
\beq
\int\limits_0^\infty d\omega \frac{\mathrm{Im}\Gamma^{\mu\nu}(\k,\k',\omega)}{\omega} = \mathrm{Re}\Gamma^{\mu\nu}(\k,\k',\omega = 0) \equiv \tilde\Gamma^{\mu\nu}(\k,\k').
\eeq
Thus, the problem of the calculation of the effective pairing interaction reduces to finding the real part of $\Gamma^{\mu\nu}$ at the zero frequency, which substantially simplifies calculations.

If we represent the order parameter $\Delta_\k$ in the form of a product of the amplitude $\Delta_0$ by the angular part $g_\k$, we can determine the dimensionless coupling parameter $\lambda$  as a result of the solution to the problem for the eigenvalues ($\lambda$) and eigenvectors ($g_\k$)~\cite{Graser}:
\beq \label{eq:lambda}
\lambda g_\k = -\sum\limits_\nu \oint\limits_\nu \frac{d\k'_{||}}{2\pi} \frac{1}{2\pi v_{F\k'}} \tilde\Gamma^{\mu\nu}(\k,\k') g_{\k'},
\eeq
where $v_{F\k}$ is the Fermi velocity, the contour integral is taken over $\k'_{||}$, belonging to the $\nu$th Fermi surface, and the band $\mu$ is unambiguously determined by the fact which of the Fermi surfaces the momentum $\k$ belongs to. The positive $\lambda$ corresponds to attraction; its maximal value corresponds to the maximal value of $T_c$,  i.e., the most favorable symmetry of the pairing and the gap function, which is determined by $g_\k$. By aligning $\lambda$  according to decreasing values, we can see what symmetries and gap structures are most favorable and which will be competing among themselves.

From the viewpoint of the mechanism of superconducting pairing, both the spin-fluctuation theories~\cite{Graser,Kemper2010,Kuroki2008} with their self-consistent generalizations in the fluctuation-exchange (FLEX) approximation~\cite{Ikeda2008,Ikeda2010,Zhang}, and the renormalization group (RG) analysis~\cite{Chubukov2008,Thomale2011} are quite complicated numerical methods. But since, in the case of pairing, it is the amplitude of scattering in the particle-particle channel on the Fermi surface that is important, the angular dependence of this amplitude can be expanded in terms of the same harmonics as
the $\Delta_\k$ is expanded. Such a method, which is called LAHA (lowest angular harmonics approximation), makes it possible to describe pairing in iron compounds both in the case of low doping and upon very strong doping with electrons or holes, using a limited set of parameters and without doing complex calculations~\cite{LAHAlong,LAHAshort,LAHAFeAs}. The main assumption of the LAHA is
the fact that the Cooper vertex $\tilde\Gamma^{\mu\nu}(\k,\k')$ can be factorized in
momenta $\k$ and $\k'$ as follows:
\beq
\tilde\Gamma^\eta(\k,\k') = \sum\limits_{m,n} C_{mn}^\eta \Psi_m^\eta(\k) \Psi_n^\eta(\k'),
\eeq
where index $\eta$ corresponds to the symmetry group of the order parameter, $C_{mn}^\eta$ are some coefficients, and the function $\Psi$ makes up the expansion in terms of angular harmonics. The expansions, depending on $\eta$, have different functional forms. Thus, for example, $\Psi_m^{A_{1g}}(\k) = a_m + b_m \cos{4 \phi_\k} + c_m \cos{8 \phi_\k} + \ldots$ for the $A_{1g}$ representation, and $\Psi_m^{B_{1g}}(\k) = a_m^* \cos{2 \phi_\k} + b_m^* \cos{6 \phi_\k} + c_m^* \cos{10 \phi_\k} + \ldots$ for the $B_{1g}$ representation.

Now, the problem can be reduced to finding a function $\tilde\Gamma_{a b}^\eta$, where $a$ and $b$ correspond to the numbers of the Fermi surfaces. For example, in Fig.~\ref{fig:spinres}, these are hole $\alpha_{1,2}$ and electron $\beta_{1,2}$ pockets. For the extended $s$ and $d_{x^2-y^2}$ wave components, we can write out the following expressions
\bea
\tilde\Gamma_{\alpha_i \alpha_j} &=& U_{\alpha_i \alpha_j} + \tilde{U}_{\alpha_i \alpha_j} \cos 2\phi_i \cos 2\phi_j, \nn\\
\tilde\Gamma_{\alpha_i \beta_1} &=& U_{\alpha_i \beta} (1 + 2 \gamma_{\alpha_i \beta} ~\cos 2\theta_1) + \tilde{U}_{\alpha_i \beta} (1 + 2{\tilde\gamma}_{\alpha_i \beta} \cos 2\theta_1) \cos 2\phi_i, \nn\\
\tilde\Gamma_{\beta_1 \beta_1} &=& U_{\beta\beta} \left[1 + 2 \gamma_{\beta\beta}(\cos 2\theta_1 + \cos 2\theta_2) + 4\gamma'_{\beta\beta} \cos 2\theta_1 \cos 2\theta_2 \right] \nn\\
&+& {\tilde U}_{\beta\beta} \left[1 + 2 {\tilde \gamma}_{\beta\beta} (\cos 2\theta_1 + \cos 2\theta_2) + 4 {\tilde \gamma}'_{\beta\beta} \cos 2\theta_1 \cos 2\theta_2 \right], \nn
\eea
where $U_{ij}$ and ${\tilde U}_{ij}$ are the interactions in the $s$ and $d$ channels, respectively, $\gamma_{\alpha_i \beta}$, ${\tilde\gamma}_{\alpha_i \beta}$, $\gamma_{\beta \beta}$, $\gamma'_{\beta\beta}$, ${\tilde \gamma}_{\beta\beta}$, ${\tilde \gamma}'_{\beta\beta}$ determine the degree of interaction anisotropy, $\phi_i$ and $\theta_i$ are the angles on the hole and electron Fermi surfaces counted from the $k_x$-axis. The equation for the order parameter is reduced here to the matrix equation $4 \times 4$, which can easily be solved. The coefficients $C_{mn}^\eta$ and all $a$, $b$, etc. entering the expansion in $\Psi$, can be obtained from a comparison with the calculation of the total $\tilde\Gamma^{\mu\nu}(\k,\k')$ using Eqs.~(\ref{eq:GammaOrb}) and~(\ref{eq:GammaBand}). A comparison of the results for the order parameters has shown that the LAHA reproduces the RPA results quite well~\cite{LAHAlong}.

\begin{figure}[t]
\includegraphics[width=\textwidth]{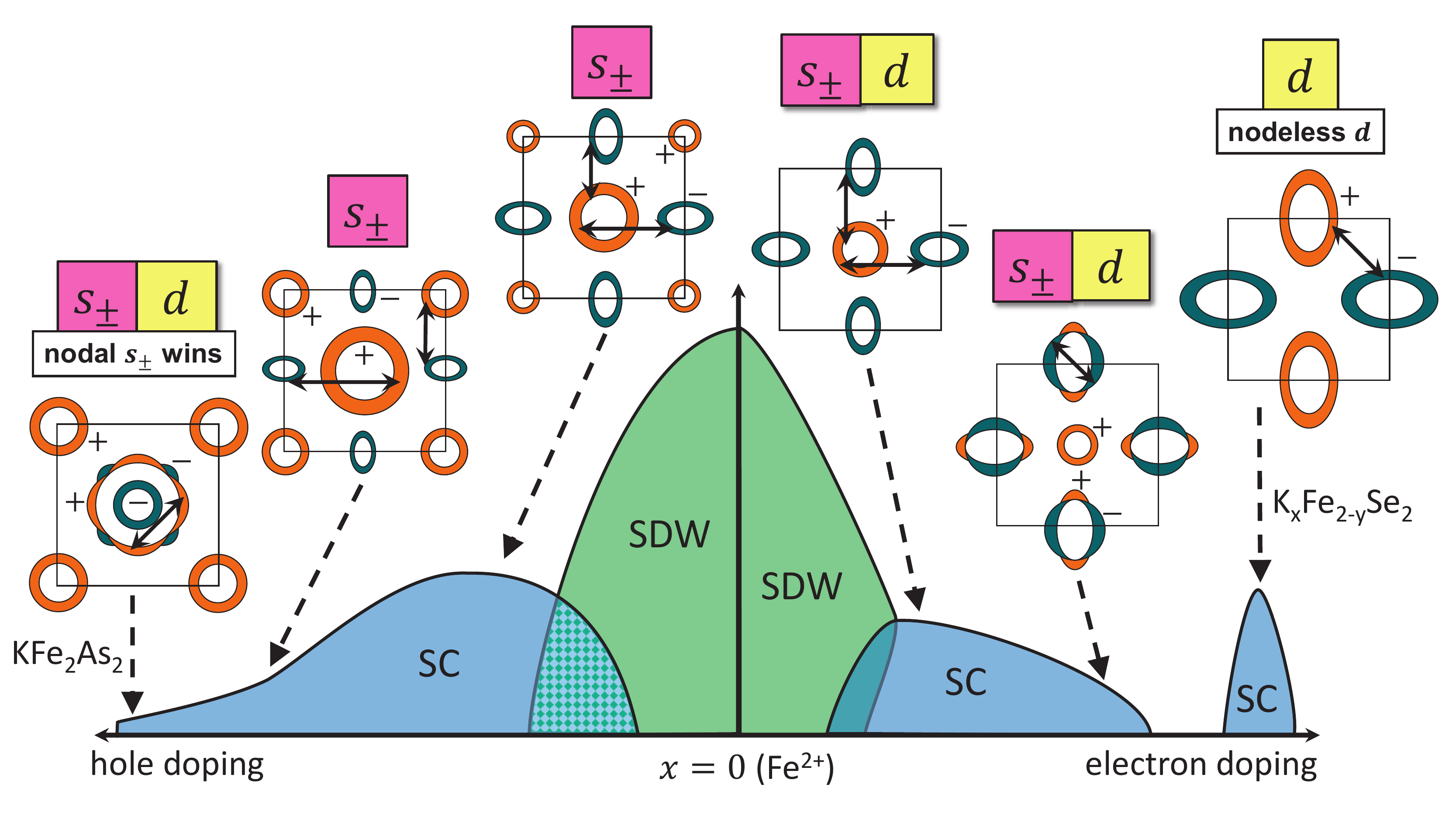}
\caption{\label{fig:phasediag} (Color online). Schematic phase diagram of iron compounds for both hole and electron dopings. The coexistence of AFM (SDW) and superconducting (SC) phases appears on a microscopic level for the case of electron doping, and on the macroscopic level (division into SDW and SC domains) upon hole doping. The qualitative picture of the symmetries of the superconducting parameter that follows from the spin-fluctuation theory~\cite{Graser,Kemper2010,review} and from the LAHA~\cite{LAHAlong,LAHAFeAs} for the two-dimensional system is shown on schematical Fermi surfaces in the insets above the phase diagram. Captions ($s_\pm$, $d$) mark the dominant and subdominant symmetries of pairing. Solid lines with an arrow at both ends ($\leftrightarrow$) indicate the dominant interaction at the Fermi surface.}
\end{figure}

One of the advantages of the LAHA is the possibility of varying the effective interaction parameters $U_{ij}$ and ${\tilde U}_{ij}$, determining thereby to which extent this or that concrete solution for the gap is stable. In this fermiological picture, one can clearly distinguish which of the interactions leads to pairing.

Fig.~\ref{fig:phasediag} schematically depicts a phase diagram and the Fermi surfaces for various levels of doping. Depending on the topology and relative volumes of the hole and electron pockets, a competition can arise between the gaps of the $s_\pm$ and $d$ types. However, it is the $s_\pm$ state that always wins in the presence of both electron and hole pockets. The dominant interactions $U_{ij}$ and ${\tilde U}_{ij}$ that were obtained from the analysis of the LAHA results are shown by arrows connecting the particles on the Fermi surfaces. Thus, the strongest interaction $U_{\alpha_i\beta}$ in the case of low doping is between the electron and hole pockets, and the dominating state has the $s_\pm$-symmetry. Upon electron doping, the repulsion $U_{\beta\beta}$ inside the electron pocket is large, and it is best for the system to form a sign-changing gap on the electron pockets in order to reduce this contribution. In this case, the $s_\pm$ state has nodal lines at electron Fermi surfaces. If the electron doping is very high (as in K$_x$Fe$_{2-y}$Se$_2$),  when the hole pockets disappear, the system forms $d$-type superconductivity because of the strong interaction between the electron Fermi pockets. One question remains open: whether such a state would be favorable as compared to the bonding-antibonding $s_\pm$ state~\cite{review,Mazin2011} upon the transformation to the Brillouin zone corresponding to two Fe atoms per unit cell. It seems that, because the spin-orbit interaction is present in this case~\cite{Korshunov2013}, and because of the
following from it hybridization along the symmetry directions, the bonding-antibonding $s_\pm$ state  should be most favorable~\cite{Khodas}. However, as follows from the calculations in the 10-orbital model for K$_{0.8}$Fe$_{1.7}$Se$_2$ and K$_{0.85}$Fe$_{1.8}$Se$_2$, it is the pairing of the $d_{x^2-y^2}$-type that always
dominates~\cite{Kreisel}.

For the hole doping, on the contrary, the appearance of a new hole pocket $\gamma$ near the point $(\pi,\pi)$ leads to the stabilization of the $s_\pm$ state without nodes on the Fermi surface. This picture is affected by the orbital character of the bands. Since the pocket $\gamma$ is mainly formed by the $d_{xy}$ orbital, as are the small regions on the electron pockets (see Fig.~\ref{fig:spinres}),  the new channel of scattering of this pocket by the electron pockets leads to the `isotropization' of the gap on electron pockets. With a further doping by holes, when the electron pockets disappear, as in KFe$_2$As$_2$, the strong interaction inside the hole pocket $\alpha_2$ forces the system to form a sign-changing gap with nodes on this pocket. The symmetry of the gap refers, as before, to the $A_{1g}$ representation and corresponds to the $s_\pm$ state with added higher angular harmonics~\cite{LAHAFeAs}.

As to the experimental observation of the $s_\pm$ state the first results were obtained via inelastic neutron scattering. Since $\chi_0(\q,\omega)$ describes the particle-hole excitations and since all excitations at frequencies less than about $2\Delta_0$ (at $T=0$) are absent in the superconducting state, the imaginary part $\mathrm{Im}\chi_0(\q,\omega)$ becomes finite only above this frequency value. The anomalous Green's functions entering Eq.~(\ref{eq.chipm}) give rise to terms proportional to $\left[1 - \frac{\Delta_\k \Delta_{\k+\q}}{E_{\k} E_{\k+\q}}\right]$.  These are the so-called \textit{anomalous coherence factors}. At the Fermi level, one has $E_{\k} \equiv \sqrt{\eps_\k^2 + \Delta_\k^2} = |\Delta_\k|$. If $\Delta_\k$ and $\Delta_{\k+\q}$ have the same sign, the coherence factors will be equal to zero, which will lead to a gradual increase in the spin susceptibility with increasing frequency in the range $\omega > \Omega_c$ with $\Omega_c = \min \left(|\Delta_\k| + |\Delta_{\k+\q}| \right)$,  whereas at frequencies lower than $\Omega_c$ we have $\mathrm{Im}\chi_0(\q,\omega) = 0$. This can be seen from Fig.~\ref{fig:spinres} for superconductivity of the classical $s$-type ($s_{++}$ state).  If, however, as in the case of $s_\pm$ and $d$ states in iron compounds, the vector $\q = \Q = (\pi,0)$ connects the Fermi surfaces with different signs of the gap, $\sgn \Delta_\k \neq \sgn \Delta_{\k+\q}$, then the coherence factors are nonzero and a jump appears in the imaginary part of $\chi_0$ at $\omega = \Omega_c$.  In accordance with the Kramers-Kronig relations, a logarithmic singularity appears
in the real part of the susceptibility. For a certain set of parameters $U$, $U'$, $J$, $J'$ entering the matrix $\hat{U}^\ppm$, the nonzero value of $\mathrm{Re}\chi_0$ and $\mathrm{Im}\chi_0 = 0$ lead to a divergence of the imaginary part of the RPA susceptibility~(\ref{eq.rpa.chippmsol}). The corresponding peak in $\mathrm{Im}\chi(\Q,\omega)$ is called the ``spin
resonance'' and appears for the frequencies $\Omega_{res} \leq \Omega_c$. This peak
is quite pronounced for the $s_\pm$ state, see Fig.~\ref{fig:spinres}. For the $d_{x^2-y^2}$ gap symmetry (although, in principle, the resonance could arise because of the sign-reversed character of the gap), the vector $\Q$ connects the states on the hole Fermi surface near the nodes of the gap $\Delta_\k$, and the total gap in $\mathrm{Im}\chi_0$, which is determined by $\Omega_c$, is very low. Since, $\Omega_c \ll \Delta_0$, the jump in $\mathrm{Im}\chi_0$ is negligibly small, and the susceptibility in the RPA shows a slight increase in comparison with that for the normal state, see Fig.~\ref{fig:spinres}. The same is true for $d_{xy}$ and $d_{x^2-y^2} + \ii d_{xy}$ gap symmetries~\cite{Korshunov2008} and for the triplet $p$-wave pairing~\cite{Maier}.

Thus, the existence of a spin resonance refers to an exclusive property of the $s_\pm$ state состояния. For iron compounds, the spin resonance was predicted theoretically~\cite{Korshunov2008,Maier}, and then revealed experimentally in the 1111, 122, and 11 families of pnictides and chalcogenides~\cite{Inosov,ChristiansonBKFA,Lumsden,ChristiansonBFCA,Park,Argyriou,Osborn,QiuFeSeTe,Babkevich}.

By introducing an additional damping of quasiparticles and by adjusting parameters, we can attain the appearance of a peak in the magnetic susceptibility in the $s_{++}$ state at frequencies above $\Omega_c$~\cite{Onari2010,Onari2011}.  From the experimental view-
point, it is important to distinguish the situation with appearing a resonance peak for $\Omega_{res} \leq \Omega_c$ from that with the enhanced susceptibility for $\omega > \Omega_c$. The first case refers to the $s_\pm$ state and indirectly confirms the spin-fluctuation mechanism of the superconductivity; the second case corresponds to the $s_{++}$ state and to the theory of superconductivity due to orbital fluctuations or electron-phonon interaction. No exact answer exists so far as to which of them is correct, but the present body of experimental data on both the spin resonance and on the quasiparticle interference scattering, penetration depth, heat capacity, and many other observed characteristics indicates in favor of the sign-changing $s_\pm$ state~\cite{review}.

Summarizing, we conclude that, in spite of the variety of the materials, the multiorbital spin-fluctuation theory of pairing can explain many observed features of iron-based superconductors, in particular, the different variants of the experimentally examined behaviors of the superconducting gap. The anisotropic $s_\pm$ state and its nodal structure on Fermi surfaces are quite sensitive to some details of the electronic
structure, such as the orbital character of the bands, spin-orbit interaction, and changes in the band structure due to the doping.

\acknowledgments

I'm grateful to A.V. Chubukov O.V. Dolgov, I.M. Eremin, P.J. Hirschfield, A. Kordyuk, I.I. Mazin, V.M. Pudalov, M.V. Sadovskii, and Yu.N. Togushova  for the useful discussions. This work was supported in part by the Russian Foundation for Basic Research (project No. 13-02-01395), by the Program No. 20.7 of the Presidium of the Russian
Academy of Sciences, by the `President Grant for Government Support of the Leading Scientific Schools of the Russian Federation' (No. NSh-2886.2014.2), by the Dynasty Foundation and the International Center for Fundamental Physics in Moscow.

\end{document}